\documentclass[baaa]{baaa}

\usepackage[pdftex]{hyperref}
\usepackage{subfigure}
\usepackage{natbib}
\usepackage{helvet,soul}
\usepackage[font=small]{caption}
\usepackage{bm}
\newcommand{\hii}{H\textsc{ii}}

\contriblanguage{1}

\contribtype{3}

\thematicarea{5}

\title{Interstellar medium and star formation}

\titlerunning{ISM and star formation}

\author{S. Paron\inst{1}
}

\authorrunning{Paron, S.}

\contact{sparon@iafe.uba.ar}

\institute{
Instituto de Astronomía y Física del Espacio, CONICET--UBA, Argentina
}

\resumen{ 
La formación de estrellas, en particular las de alta masa, posee varios interrogantes
aún abiertos. Hoy en día, gracias a los telescopios e instrumentos más modernos,
estamos logrando poder observar y analizar los procesos físicos y químicos
involucrados en el nacimiento de las estrellas masivas.
En este trabajo se introduce al medio interestelar, cuna de las estrellas,
haciendo foco en las estructuras interestelares distribuidas en las
distintas escalas espaciales relacionadas al colapso del gas que da lugar a los
procesos de formación estelar.
A través de algunos trabajos actuales del grupo de investigación de Medio Interestelar, Formación Estelar y Astroquímica 
del Instituto de Astronomía y Física del Espacio (\href{https://interestelariafe.wixsite.com/mediointerestelar}{https://interestelariafe.wixsite.com/mediointerestelar}), se muestra que el estudio
observacional de la formación de estrellas es una investigación que debe realizarse
de manera multiespectral apuntando a la multiescala espacial. 
}

\abstract{
The formation of stars, particularly the high-mass star formation, poses several still open questions. Nowadays, 
thanks to the most modern telescopes and instruments,
we are able to observe and analyse many physical and chemical processes
involved in the birth of massive stars.
This work introduces to the interstellar medium, cradle of the stars,
and makes focus on the interstellar structures distributed in the
different spatial scales related to the collapse of the gas that gives rise to the
star formation processes.
Through some current works done by the investigation group of Interstellar Medium, Star Formation and Astrochemistry belonging to
Instituto de Astronomía y Física del Espacio (\href{https://interestelariafe.wixsite.com/mediointerestelar}{https://interestelariafe.wixsite.com/mediointerestelar}), it is shown that the 
observational study of the star formation is a research that must be carried out
in a multispectral way, pointing to the spatial multiscale.

}


\begin{document}

\maketitle

\section{Introduction}

\begin{figure}[h!]
\centering
\includegraphics[width=9cm]{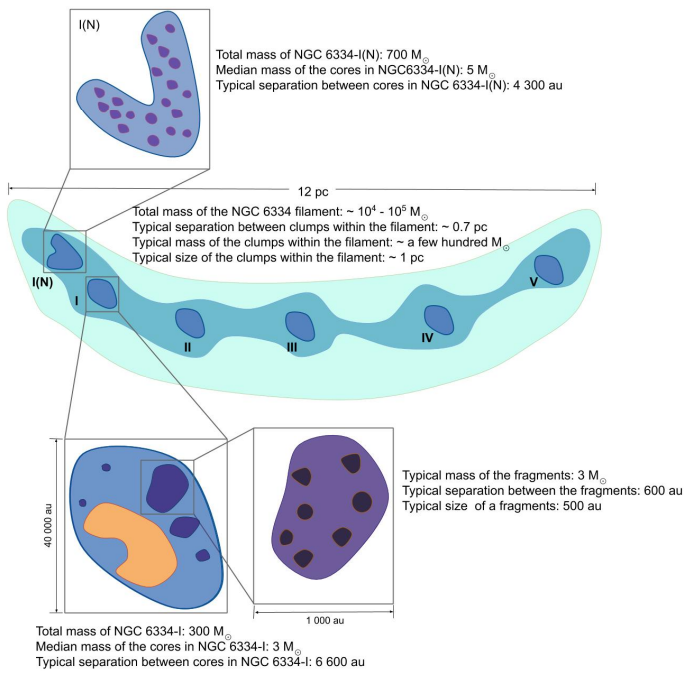}
\caption{Schematic view of a filament in NGC\,6334 with its hierarchically fragmentation into clumps and cores. Image extracted from \citet{sada20}. }
\label{hier}
\end{figure}

Nowadays we know that at the large spatial scales, the interstellar medium (ISM) is composed by an intricate filamentary network of gas and dust \citep{andre14,pineda23}. Such filaments can have tens of parsecs in large, with different sizes in width, having, the thinnest ones, a median value around 0.1 pc \citep{arzou19}. The filaments may hierarchically fragment into clumps and cores, i.e. structures that goes from parscecs down to 0.1 pc and hundreds of astronomical units (au). To illustrate it, Figure\,\ref{hier}, extracted from \citet{sada20}, shows a schematic view of a filamentary cloud in NGC\,6434, in which sizes and masses of the molecular features at different spatial scales are included.  

The molecular gas in the ISM, distributed in clouds, that nowadays are resolved into filaments, is in constant interaction with the radiation from the stars, particularly from the high-mass ones, and with expansive events such as those produced by \hii~regions and supernova explosions. Such events generate bubbles, and usually the compression of the gas due to their expansion produces new filament-like structures \citep{pineda23}. 

\begin{figure}[h]
\centering
\includegraphics[width=8.5cm]{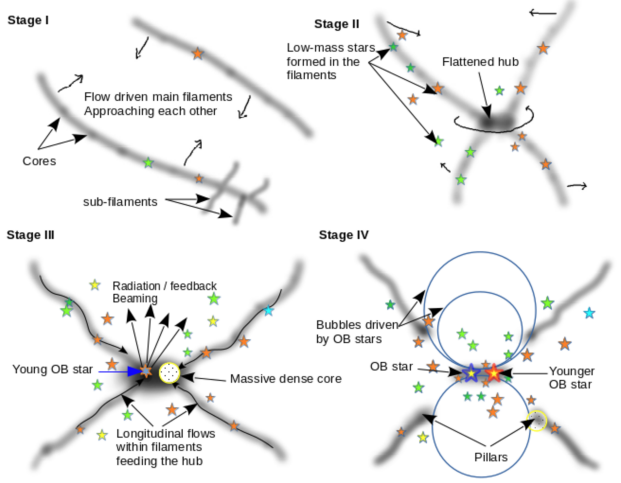}
\caption{Sketch showing the different stages proposed as part of the ``filaments to clusters'' unified paradigm for star formation. 
Image extracted from \citet{kumar20}. }
\label{fil}
\end{figure}

The filaments are not statics, they have their own dynamics and they can interact between them, converging and overlapping in junctions called as hubs \citep{kumar20}. Figure\,\ref{fil} shows different stages in this dynamics that in turn is related to the star formation processes. In other words, we can point out that the stars form in cores, embedded in clumps, that are embedded in the interstellar filaments. In that sense, a new paradigm seems to emerge, this is a ``filaments to clusters'' unified paradigm for star formation \citep{kumar20}. 

In this paradigm, \citet{kumar20} proposed that the
low-intermediate-mass stars form slowly (10$^{6}$ yr) in the filaments, and massive stars form quickly (10$^{5}$ yr) in the hubs.
The sketch presented in Fig.\,\ref{fil} shows that the hubs are formed due to the overlapping of flow-driven main filaments (Stage\,I). Then, such hubs gain a
twist as the overlap point is different from the centre of mass, and the density is enhanced due to the addition of filament densities (Stage\,II). Low-mass stars
can be ongoing before and during hub formation. The hub gravitational potential triggers and drives longitudinal flows that brings more matter
and the density of the region enhances. 
Once massive stars are formed within the hubs after its fragmentation, radiation pressure and ionisation feedback
escapes through the inter-filamentary cavities by generating holes in the hub (Stage\,III). Finally, the expanding radiation bubbles create \hii~regions, 
breaking out the filaments and producing gaseous interstellar features known as pillars (Stage\,IV). The result of these processes is a
mass-segregated embedded cluster, with a mass function that is the sum of stars continuously created in the filaments and the massive stars formed in the hub \citep{kumar20}.

Regarding the processes that take place at smaller spatial scales, it is important to take into account the fragmentation of the mentioned filaments, and mainly, 
the fragmentation of the embedded molecular clumps into molecular cores. 

It is well known that the formation of the stars, particularly of the high-mass ones, begins with the fragmentation of a massive clump into molecular cores. However, it is still unknown whether this fragmentation gives rise to prestellar cores massive enough (a few tens of solar masses) to form directly this type of stars, or leads to cores of low and intermediate masses that acquiring material from their environment can generate massive stars \citep{palau2018,moscadelli2021}. In the first scenario, it is proposed that high-mass stars form through an individual monolithic core collapse, and in the second one, they form from a global hierarchical collapse of a massive clump, where many low and intermediate mass cores competitively accrete material from the surrounding through converging gas filaments that feed the cores \citep{motte18,sch19,vazquez2019}. Nowadays many efforts are being made to unravel which model of high-mass star formation best explains the born of such a type of stars.

\begin{figure*}[h]
\centering
\includegraphics[width=16cm]{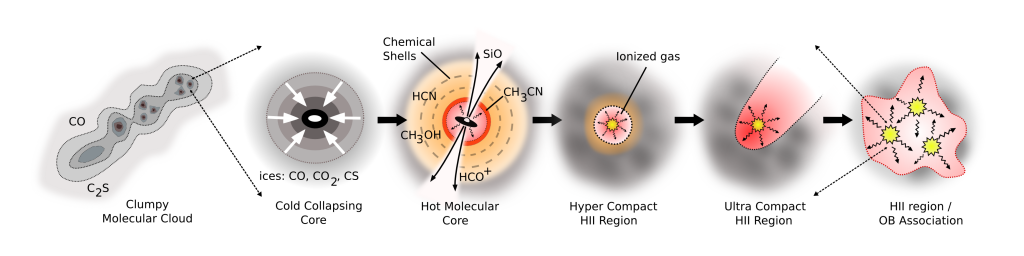}
\caption{Sketch showing the evolutionary sequence for high-mass stars formation.
Credits: Cormac Purcell.}
\label{seq}
\end{figure*}

Hence, taking into account that all processes that occurs at the larger spatial scales have consequences in the smaller ones, and vice versa, to advance in our general knowledge about star formation, it is important to study the ISM in a comprehensive way.
For instance, Figure\,\ref{seq} displays the evolutionary sequence for high-mass stars formation, in which all the ``intrinsic'' processes, such as gas collapse and accretion, outflows, ionisation and photodissociation, have a deep impact in the surroundings. 

As Figure\,\ref{seq} shows, molecules are ubiquitous along the different stages of star formation, even in the \hii~region stage. The detection of molecular species, many of them very complex, organic and pre-biotic, can be used to trace and to probe the involved physical processes. Certainly, the molecular cores with active star formation, known as hot molecular cores, are the chemically richest regions in the ISM (e.g. \citealt{herbst09,bonfand19}). The star forming processes strongly influence the chemistry of such cores and their surroundings \citep{jorgen20}. Hence, observing molecular lines, and studying their emission and chemistry, is very useful to characterise the physical and chemical conditions of the gas. This is indeed an important issue to trace the different stages in the fragmentation of the gaseous structures and the subsequent star formation. 

In this article, I review some recent works done in our research group putting them in a general context regarding the investigation of the ISM. I present them separated into results obtained from large, intermediate, and small spatial scales.

\section{Large spatial scales: photodissociation and fragmentation} 

The large structures of the ISM, the filaments and clouds, have dusty and gaseous fragments within them, smaller features known as molecular clumps and cores that can lead to the formation of the stars. Beside the gravity and the turbulence, the radiation, and in particular the far ultraviolet (FUV) photons that photodissociate the molecular gas, can play an important role in the fragmentation of different interstellar molecular features. 

Given that it is not possible to directly measure the FUV radiation toward distant, large and obscured regions, we can evaluate the action of such photons through their consequences in the molecular gas. 

It is known that the C$^{18}$O is selectively photodissociated by the FUV radiation with respect to the $^{13}$CO. This phenomenon was observed in 
many relatively nearby molecular clouds that are exposed in different ways to the radiation field \citep{yama19,kong15,minchin95}. The selective photodissociation of the C$^{18}$O is evaluated from the $^{13}$CO/C$^{18}$O ratio (X$^{13/18}$), and it was analysed in clouds that the sources responsible of the FUV photons are embedded OB stars \citep{shima14}, and in clouds that are only affected by the interstellar radiation field \citep{lin16}.

\begin{figure}[h!]
\centering
\includegraphics[width=8.5cm]{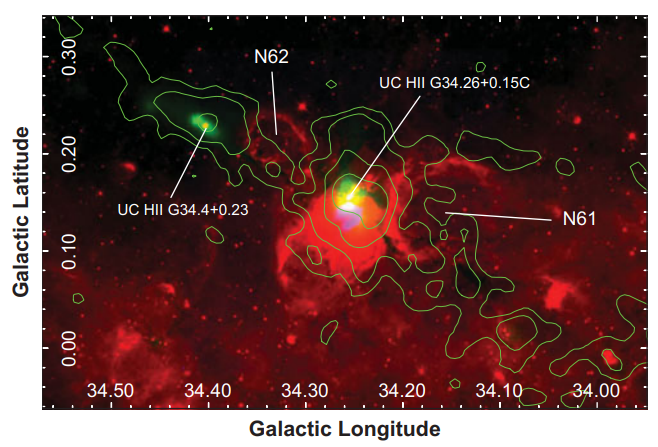}
\caption{Three-colour image of the IRDC 34.43+0.24 in which the Spitzer-IRAC 8 $\mu$m
is displayed in red, the radio continuum emission at 20 cm in blue, and the continuum emission at 1.1 mm in green. The green contours represent the C$^{18}$O J=1--0 integrated emission. Image extracted from \citet{areal19}.}
\label{irdc34}
\end{figure}

Using the X$^{13/18}$~ratio as a tool to evaluate the molecular gas photodissociation, we studied several star-forming regions, and we pointed out that the fragmentation of the molecular gas in such regions are not only produced by gravity and turbulence, but also for this phenomenon.  For instance, in \citet{areal19} we studied in deep the filamentary infrared dark cloud IRDC 34.43+0.24 located at a distance of about 4 kpc. Figure\,\ref{irdc34} presents this region, showing the presence of a filament like molecular feature (traced by the C$^{18}$O emission) that it is affected by two extended \hii~regions (N61 and N62), and additionally there are two embedded ultracompact \hii~regions. The molecular gas in the C$^{18}$O emission appears clumpy and quite fragmented, being interesting to study the influence of the FUV radiation from both, the observed \hii~regions and the ISM radiation field. 

\begin{figure}[h]
\centering
\includegraphics[width=8cm]{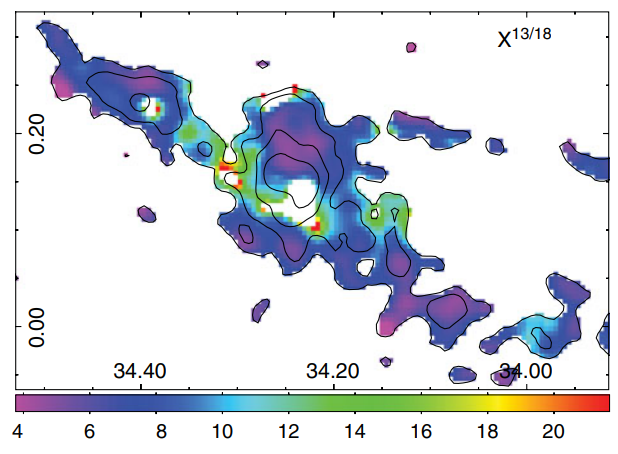}
\caption{X$^{13/18}$ ratio obtained toward IRDC G34.43+0.24. The contours represent the C$^{18}$O J=1--0 integrated emission as presented in Fig.\,\ref{irdc34}. Image extracted from \citet{areal19}.}
\label{xirdc34}
\end{figure}

After calculating pixel by pixel the $^{13}$CO and C$^{18}$O column densities (N$^{13}$ and N$^{18}$) using the $^{12}$CO, $^{13}$CO and C$^{18}$O J=1--0 line, and assuming local thermodynamical equilibrium (LTE), the X$^{13/18}$ was obtained from the N$^{13}$/N$^{18}$ ratio. Figure\,\ref{xirdc34} presents a map of 
X$^{13/18}$ toward IRDC G34.43+0.24. It can be noticed some regions with high X$^{13/18}$ values, showing sectors in the filament feature in which the FUV photons are penetrating and destroying the molecular gas. In general, they are located close to the \hii~regions, being remarkable the thinner portion of the filament, suggesting that in a future such a filament will be broken at this position producing at least two large fragments, which in turn could have sub-fragments. On the other hand, there are several regions with low X$^{13/18}$ values (violet colour in Fig.\,\ref{xirdc34}) indicating that they are shielded from the FUV. In general, such regions coincide with dark clouds and high density structures. Blanked pixels in the figure are due to optical depths effects in the $^{13}$CO emission that prevent to obtain reliable results toward these regions.

\begin{figure}[h]
\centering
\includegraphics[width=8cm]{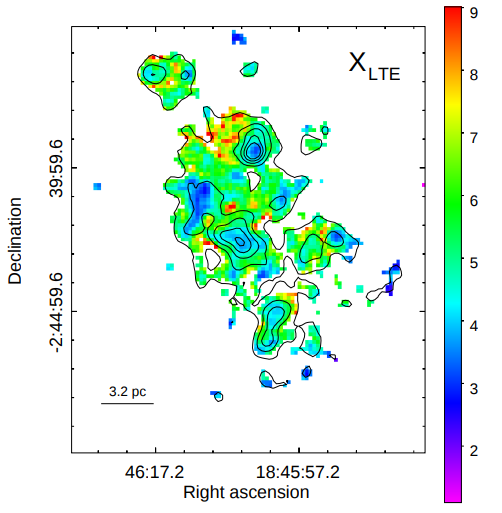}
\caption{X$^{13/18}$ ratio obtained toward G29.96-0.02 from the J=3--2 of the CO isotopes using LTE considerations. The contours represent the C$^{18}$O J=3--2 integrated emission delimiting the molecular cloud/filament analysed. Image extracted from \citet{paron18}.}
\label{xg29}
\end{figure}

In another work, the J=2--1 and J=3--2 lines of the carbon monoxide isotopes were used to perform a similar analysis toward the massive star forming region G29.96-0.02 \citep{paron18}. In this case, LTE and non-LTE analysis were done toward this region located at a distance of about 6.2 kpc. For instance, here I present in Figure\,\ref{xg29}, the X$^{13/18}$ ratio obtained by assuming LTE
toward G29.96-0.02. As in the former case, this analysis is useful to study the molecular gas at large spatial scales (see the scale bar in Fig.\,\ref{xg29}) identifying regions that will be fragmented due to the FUV action (green, yellow and red colours) and dense regions in which the radiation is staling (cyan and blue colours). 

It is worth noting that the difference in the values of the X$^{13/18}$ ratios obtained in the G29.96-0.02 and IRDC G34.43+0.24 could be due to their different distances respect to the galactic centre \citep{wilson94}. Additionally, as it was suggested in our first work about selective photodissociation done toward 114 sources \citep{areal18}, these differences must be analysed focusing on the type of sources that are embedded, or located close to, the molecular clouds/filaments. Finally, it is important to remark that the use of different CO rotational transitions to derive the X$^{13/18}$ ratios is an issue that deserves more analysis. This is still open in our investigations. 

In conclusion, this kind of investigations contribute to understand fragmentation processes and molecular gas destruction at large spatial scales, useful to study, in a next step, processes occurring within the clouds and filaments.

\section{Molecular clumps, intermediate spatial scales}

Molecular clumps are dense fragments embedded in the interstellar filaments. Regarding the comprehensive study of the star-forming processes, and the related conditions of the 
ISM, such clumps represent the next step, after the filaments, to investigate the interstellar chemistry and physics. 
For instance, massive young stellar objects (MYSOs) generate energetic molecular outflows (e.g. \citealt{arce07}) that 
are produced by jets/winds generated by the protostars and accretion discs. They contribute to the removal of excess of
angular momentum from accreted matter and to the dispersal of infalling circumstellar envelopes at the clump spatial scales. 
Therefore, even though the molecular outflows usually may be confused with the molecular gas of the clump in which
the MYSOs are embedded, it is worth making efforts in studying them. 

Protostellar winds, jets, and hence, the extended molecular outflows, produce a deep impact in the chemistry of both, cores and molecular clumps \citep{arce07,jorgen20,rojas24}. Thus, for a complete understanding of the star-forming processes, it is necessary to study such physical processes together with the related chemistry.  

\begin{figure}[h]
\centering
\includegraphics[width=6.7cm]{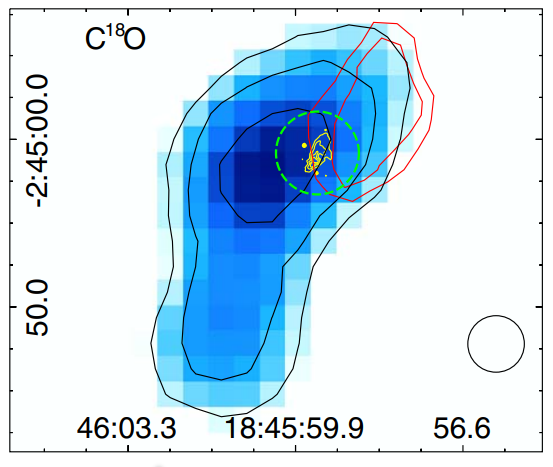}
\caption{Integrated C$^{18}$O J=3--2 emission displayed in blue with black contours of the
molecular clump in which YSO\,G29 is embedded. 
The red contour corresponds to the $^{12}$CO J=3--2 emission representing a redshifted molecular outflow. The beam of such observations is at the bottom right corner. 
The dashed circle is the pointing and the beam of the ASTE observations. The yellow contours represent the near-IR emission of YSO\,G29 (see Sect.\,\ref{small} and Fig.\,\ref{g29niri}). Image extracted from \citet{areal20}.}
\label{clumpg29}
\end{figure}

Southward the massive star-forming region G29.96-0.02 presented above (see Fig.\,\ref{xg29}), a YSO lies in a high-density region shielded from the FUV radiation. 
In \citet{areal20}, we analysed the molecular clump in which such a YSO (hereafter YSO\,G29) is embedded. Figure\,\ref{clumpg29} displays this clump in the C$^{18}$O emission and the discovered molecular outflow (red contours). Such an outflow is gas moving away from us, i.e. a redshifted molecular outflow, and its blue counterpart was not observed because it is confused with the southward molecular gas of the clump. This analysis was done with data from the James Clerk Maxwell Telescope (JCMT) retrieved from its database. Additionally, we observed some molecular species (HCN, HNC, HCO$^{+}$, and C$_{2}$H) towards the position of YSO\,G29 (green dashed circle in Fig.\,\ref{clumpg29}) using the Atacama Submillimeter Telescope Experiment (ASTE). The detection of HCO$^{+}$
is consistent with the presence of jets and flows of gas (e.g. \citealt{raw04,sanchez13}) confirming the presence of molecular outflows in the region. The other molecules were useful to
prove the presence of dense gas in the clump and to estimate the kinetic temperature in T$_{\rm K}\sim23$ K, which is an useful parameter
to characterise the conditions of the external gaseous layers of the clump in which a YSO is embedded. 

\begin{figure}[h]
\centering
\includegraphics[width=7cm]{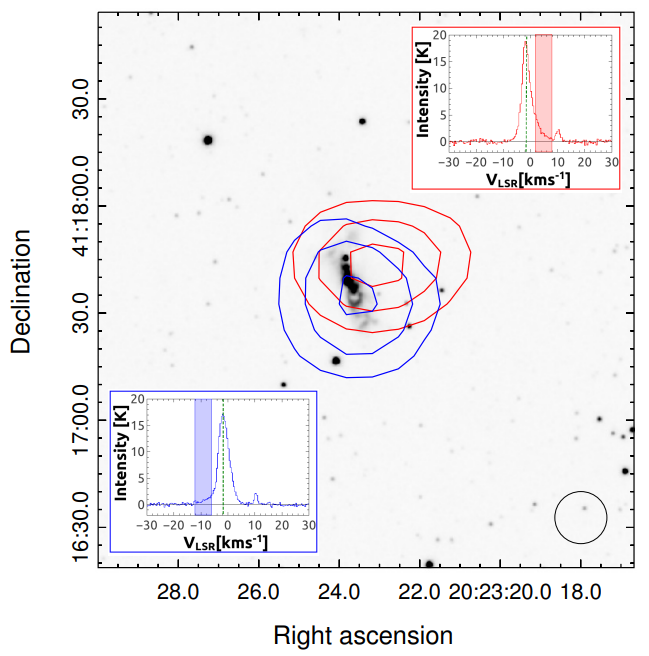}
\caption{
Near-IR image from UKIDSS at the Ks band of YSO\,G79. The blue and red
contours represent the integrated $^{12}$CO J=3--2 emission (JCMT data) in two velocities
ranges that are the spectral wings (shown in the spectra included in the inset images). The JCMT beam of 14.5\arcsec~is included at the bottom right corner. Image extracted from \citet{paron22}.
}
\label{clumpg79}
\end{figure}

In a more recent work, a quite isolated and nearby (about 1.4 kpc) MYSO was studied (YSO\,G079.1272+02.2782; hereafter
YSO\,G79) at different spatial scales \citep{paron22}. As done in the case of YSO\,G29, we 
looked for evidence of outflow activity, finding two CO lobes, one redshifted and the other one
blueshifted, in coincidence with an intriguing near-IR emission emanating from this YSO (see Fig.\,\ref{clumpg79}). This analysis was useful to generate a better characterisation of such molecular outflows that were not correctly catalogued in the literature, and it motivated us to perform a chemical analysis of the region at the clump spatial scales. 

Using in this case data from the database of the Pico Veleta IRAM Telescope obtained from \citet{gerner14}, we identified several molecular lines toward the molecular clump in which YSO\,G79 is embedded. The analysis of HCN, HNC, H$^{13}$CO$^{+}$, HC$_{3}$N, H$_{2}$CO, among others, allowed us to estimate a temperature of about 20 K for the external layers of the clump, and to uncover a complex chemistry in them (chemistry at clump scales), finding that the inner jets (see Sect.\,\ref{small}) does not have any influence on such a chemistry yet.

In conclusion, besides the study of the physical processes that take place within the molecular clumps, the study of the chemical processes are crucial to complement such physical investigations, and usually, the chemistry can be used to trace evolutive stages. In this line I would like to mention a very recently work done in our group in 
which some chemical tools were probed toward 55 sources representing MYSOs at different evolutionary stages: infrared dark clouds (IRDCs), high-mass protostellar objects (HMPOs), hot molecular cores (HMCs) and ultracompact \hii~regions. From this large sample, \citet{martinez24}  
found that the use of the HCN/HNC ratio as a universal thermometer in the ISM as it was proposed in the literature should be taken with care, and the HCN optical depth effects must be considered. This tool should be utilised only after a careful analysis of the HCN spectrum, as it was done in the above mentioned works in which the temperature was estimated through this method.  Additionally, \citet{martinez24} point out that the emission of H$^{13}$CO$^{+}$, HC$_{3}$N, N$_{2}$H$^{+}$ and C$_{2}$H could be useful to trace and distinguish regions among IRDCs, HMPOs and HMCs, 
supporting that these molecular species can be used as ``chemical clocks''.

\section{Analysing the small spatial scales} 
\label{small}

At the smaller spatial scales ($\lesssim$0.1 pc), meaning the typical molecular cores (and even smaller) scales, there are structures that 
represent the deepest ``pieces'' of ISM that deserve very detailed investigations in order to understand how stars are born. In the case of high-mass stars,
the analysis of this last stage of fragmentation is useful to discern what model of formation is more appropriated: monolithic collapse or competitive accretion (e.g. \citealt{motte18,palau2018,moscadelli2021}).

Given that these molecular features are small and they are deeply embedded in very dark regions, radio and millimeter wavelengths interferometers must mainly be used to observe and analyse them. The Atacama Large Millimeter Array (ALMA) is the most appropriated instrument to perform this kind of investigations, and from a multiwavelength perspective, their high-sensitivity and high-angular resolution data can be combined with, for instance, infrared data to obtain the most complete picture possible of the star-forming processes in these regions.

\begin{figure*}[h!]
\centering
\includegraphics[width=14cm]{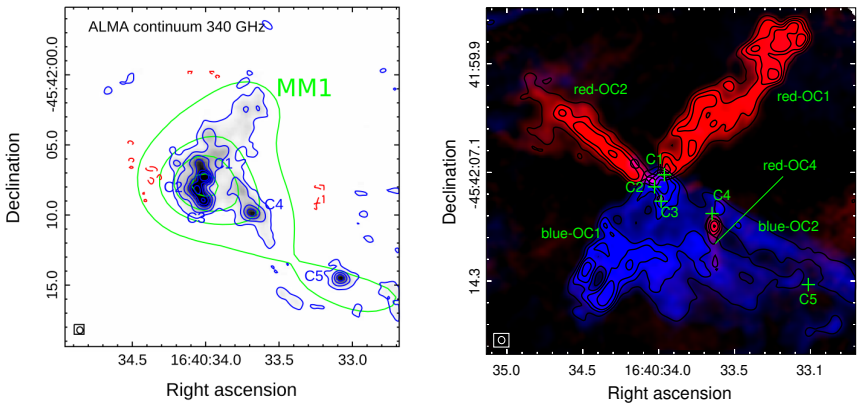}
\caption{
Left: ALMA continuum emission at 340 GHz (12\,m array) in grayscale with blue contours. The
green contours correspond to the ALMA continuum emission at 340 GHz (7\,m array).
Right: Molecular outflows obtained from the integration of the spectral red and blue shifted wings in the $^{12}$CO J=3--2 line. Images extracted from \citet{ortega23}.
}
\label{g338}
\end{figure*}

In a recent work \citep{ortega23}, we investigate the fragmentation taking place in the massive
clump AGAL\,G338.9188+0.5494. We discover that a massive molecular clump surveyed with the Atacama Pathfinder Expermient (APEX) Bolometer Array LABOCA at 870 $\mu$m \citep{cseng14}, is fragmented in two structures if the region is analysed with ALMA data obtained from the 7\,m array configuration  (4\arcsec~of angular resolution; see green contours in Fig.\,\ref{g338}-left). However, if we analyse ALMA data obtained from the 12\,m array configuration (about 0.5\arcsec~in angular resolution) such a molecular feature is resolved at least into five cores (see map and blue contours in Fig.\,\ref{g338}-left). This study allowed us to characterise such molecular cores, and using $^{12}$CO emission from the same ALMA data set, to discover a complex of molecular outflows (see Fig.\,\ref{g338}-right) indicating that some cores are already active (i.e. they are hot molecular cores with protostars in their interiors). This finding is in agreement with what is observed at the 4.5 $\mu$m emission from Spitzer observations, that based on them, the source was catalogued as an extended green object (EGO; \citealt{cyga2008}). An EGO is a potential source of energetic molecular outflows. Additionally, the near-IR emission at the Ks-band from the VISTA survey (see Figure\,4 in \citealt{ortega23}) supports the presence of jets and outflows in the region. Regarding the high-mass star formation models, our analysis suggests that competitive accretion is the more likely process in this region. 

\begin{figure}[h!]
\centering
\includegraphics[width=6.5cm]{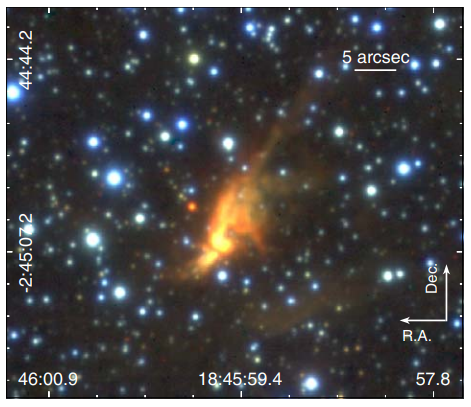}
\caption{Three colour image of the JHKs broad bands obtained with NIRI at Gemini North toward YSO\,G29. Image extracted from \citet{areal20}.
}
\label{g29niri}
\end{figure}

At these spatial scales, the near-IR data at high angular resolutions, for instance data from the Gemini Telescopes instruments, are very useful to study the inner processes that take place close to the protostars. In that sense, following our multispatial scales and multiwavelength studies, YSOs\,G29 and G79 from the regions presented above were observed with Gemini, using NIRI and NIFS instruments, respectively.

YSO\,G29 was observed with NIRI at the JHKs broad bands (see Fig.\,\ref{g29niri}) among other photometric observations of different lines. Intriguing features were found suggesting a disk-jets system. Northern jet-like structures were related to what was found at the larger spatial scale (see above), particularly to the redshifted molecular outflow, while the smaller southern jet-like feature was explained based on the region of the molecular clump where the YSO is embedded \citep{areal20}. However this scenario is not completely understood, and new observations at radio continuum from the Very Large Array (VLA) and data from ALMA are being analysed (Martinez, N.C., et al., in prep.). 

\begin{figure}[h]
\centering
\includegraphics[width=6.3cm]{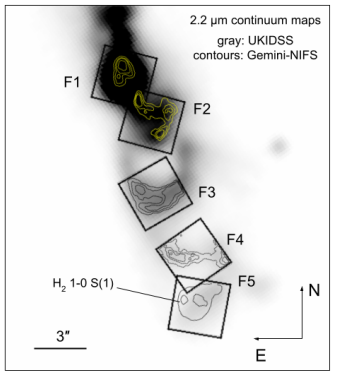}
\caption{Ks-band emission (in gray) obtained from UKIDSS. The contours in each field are the continuum at 2.2 $\mu$m obtained from the NIFS-Gemini observations, except in F5 where some contours of the H$_{2}$ S(1) 1--0 emission are displayed.  Image extracted from \citet{paron22}.
}
\label{g79nifs}
\end{figure}

In the case of YSO\,G79, near-IR integral field spectroscopic observations were carried out using NIFS at Gemini North towards five fields (F1--F5) along a jet-like structure associated with the molecular outflows presented above (see Fig.\,\ref{g79nifs}). The analysed continuum emission and the several H$_{2}$ lines and the Br$\gamma$ emission allowed us to suggest a possible binary system of protostars generating a precessing jet \citep{paron22}. These observations resolve structures of about 300 au, and by relating such a jet to the outflows found at the larger spatial scale, we proposed that we are observing the consequences in the surrounding material of a blueshifted jet generating helical flows. A bow shock coming to us, and seen almost from its front (see the H$_{2}$ emission contours in F5 in Fig.\,\ref{g79nifs}), was analysed morphological and kinematically.

\section{Astrochemistry}

As seen along the article, molecules give us important information about the physical processes that take place in the dusty and gaseous structures at all spatial scales. To use such molecules as tools to investigate the ISM it is necessary to understand its chemistry: formation and destruction reactions, which in turn can give us information about the physical conditions of the regions in which they take place.  

Even though chemistry occurs along all spatial scales in the ISM, its investigation is particularly interesting at the smallest ones, i.e. at core scales. This is because when the temperature increases in such cores, becoming them in hot cores, many complex organic molecules desorbe from the dust surfaces generating a very rich chemistry (e.g. \citealt{busch22}). 

\begin{figure*}[h]
\centering
\includegraphics[width=12.6cm]{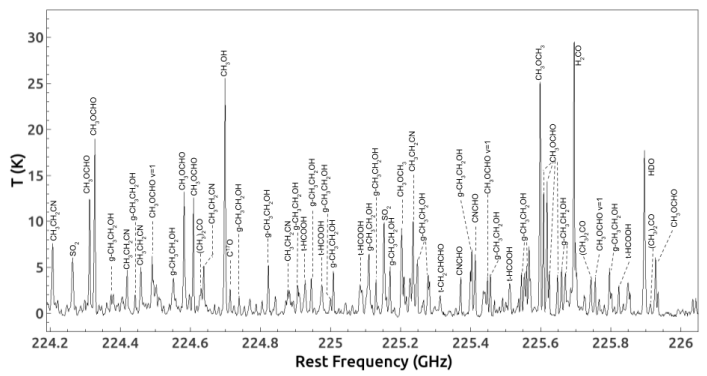}
\caption{Molecular lines identified in the spectral window spw3 of ALMA data in Band\,6 toward core\,1 (C1) in the G338.9188+0.5494 region (see Fig.\,\ref{g338}). Image extracted from \citet{ortega23}.
}
\label{espectg338}
\end{figure*}

For example, in \citet{ortega23} we performed an astrochemical study of a star-forming region, in particular of the more active core (core\,1) in 
the clump G338.9188+0.5494 (see Fig.\,\ref{g338}). This allowed us to complement the physical studies of the core obtaining important information, not only in what respect to the star formation models, but also about the chemical conditions that encompass to the star-forming processes. As an example of the chemical richness of such a core, I present here just some molecular lines detected toward it (see Fig.\,\ref{espectg338}).

\begin{figure}[h]
\centering
\includegraphics[width=6.6cm]{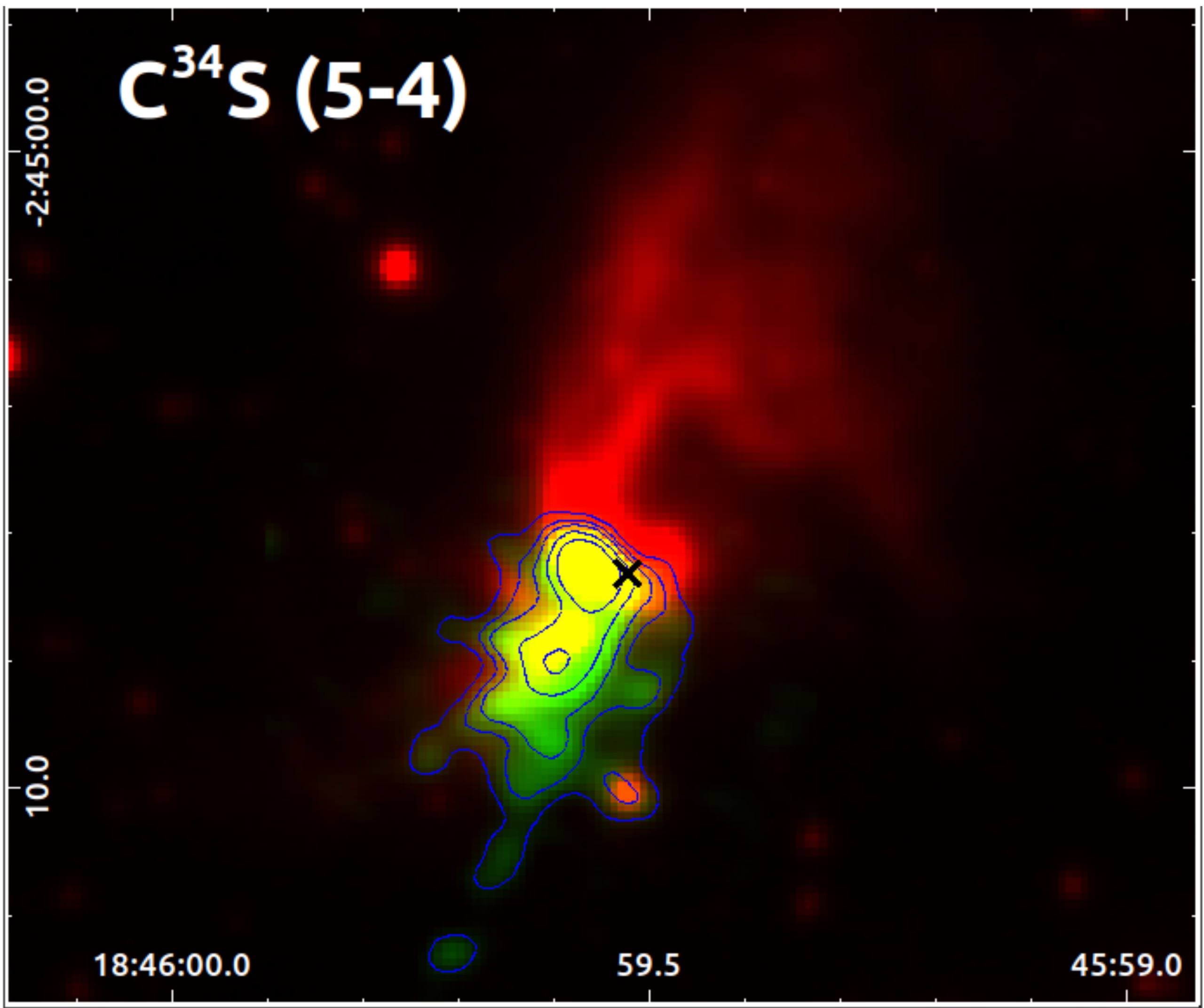}
\caption{Integrated emission of the C$^{34}$S (5--4) line from ALMA data (green with contours) over the Ks emission from NIRI Gemini (see Fig.\,\ref{g29niri}). The black cross indicates the position of a compact radio continuum source at 3 cm detected with VLA (Martinez, N.C., et al. in prep.). 
}
\label{CSg29}
\end{figure}

Another astrochemical work that I would like to mention here is the identification of molecular species toward YSO\,G29, that together with new Gemini NIFS and VLA data, will be useful to better understand the scenario of this source, following in this way, an investigation that began several years ago covering all spatial scales (Martinez, N.C., et al. in prep.). Emission lines of the identified molecular species toward YSO\,G29 are being integrated with the aim of obtaining chemical information from maps that display the morphology of the molecular emissions. For an example, I just present here the integrated emission of the C$^{34}$S (5--4) line (Fig.\,\ref{CSg29}). A preliminary work about this study appears in this number of the BAAA (see Martinez, N.C., et al.). C$^{34}$S is a sulphur-bearing molecule that can give us important information about the conditions of the dense gas. In this particular case, as it is seen in the figure, such a dense gas is mainly distributed toward the south of YSO\,G29, suggesting that the southern jet-like feature as seen at near-IR (see Fig.\,\ref{g29niri}) is small due to the encounter of the jet with high-density gas. By the way, the study of the sulphur-bearing chemistry is a matter of great interest in current astrochemical investigations \citep{fontani23}.

Additionally, it is worth mentioning that top-symmetric molecules such as methyl cyanide (CH$_3$CN) and methyl acetylene (CH$_3$CCH) are usually used as temperature tracers. Understanding their chemistry and the limitations of the temperature estimate methods is important to study star-forming regions. The investigations of these molecules are also being developed in YSO\,G29, and toward a large sample of hot molecular cores (Marinelli, A., et al. in prep.). Preliminary results about this last work appears also in this number of the BAAA (see Marinelli, A., et al.).

\section{Conclusion and gratitude} 

In these last lines I include only one point as a scientific conclusion. And also as an important general conclusion I would like to express my gratitude to the job, effort and accompaniment of some persons that work and have worked together with me in these investigations. 

As a science conclusion I mention that the ISM is a very complex, active and dynamic medium. As the stars and planets form from matter of such a medium, studying it is relevant to understand our own origins, and the origins of others exotics stars and planetary systems. The study of the ISM, and in particular of the star formation, must be done through observations and computational models, and the observations must be multiwavelength. As I pointed out here, to have a comprehensive understanding of the processes that take place in the ISM, we must study all spatial scales. 

Finally, I consider an important conclusion expressing my gratitude to some partners that work and have worked with me, that contribute and have contributed to form our research group \href{https://interestelariafe.wixsite.com/mediointerestelar}{(link)}.  
First of all, I express my gratitude to my lifelong partner,
Dr.\,Martín Ortega, with whom we have built this research group together learning to work in science in a different way.
Then, to the PhD student Lic.\,Naila Martínez, graduate
in chemical sciences, with whom we find ourselves in the challenge
to open a new line of research in the group: the astrochemistry.
In addition, I want to say that both of them were a big support to me in a special personal and professional moment.

I also want to mention two persons who have obtained PhD degrees
working in the group: Dra.\,Mariela Celis Peña, with whom we not only stayed
with the ISM of our galaxy, we also `went to' the Large Magellanic Cloud, learning many new things. She left in our group a nice human warmth and her love for science. Dra.\,María Belén Areal, with whom, thanks to her great dedication in the
research work, we have learnt a lot, laying some important foundations for the present and future investigations that we carry out. 
They indeed have contributed to the 
science with excellent PhD thesis (see \href{https://bibliotecadigital.exactas.uba.ar/collection/tesis/document/tesis_n6906_CelisPena}{MCP}, and \href{https://bibliotecadigital.exactas.uba.ar/collection/tesis/document/tesis_n7240_Areal}{MBA}).

The science and a research group are not built from one day to another; like the stars, it requires many processes over a long time, and these depend on the people who contribute to the task. To all the people who build and dream, my sincere gratitude.

\begin{acknowledgement}
I acknowledge support from CONICET and ANPCYT through grants PIP 2021 11220200100012 and PICT 2021-GRF-TII-00061. I thank to the SOC and LOC of the AAA meeting for the invitation and the support received. 
\end{acknowledgement}


\bibliographystyle{baaa}
\small
\bibliography{refparon}
 
\end{document}